\documentclass[aip,jmp,amsmath,amssymb,reprint]{revtex4-1}
\usepackage{graphicx}
\usepackage{dcolumn}
\usepackage{bm}
\usepackage{hyperref}

\begin{document}

\title{Energy distribution and quantum yield for photoemission from air-contaminated gold surfaces under UV illumination close to the threshold}
\author{Gerald Hechenblaikner}
\email{Gerald.Hechenblaikner(at)astrium.eads.net}
\affiliation{EADS Astrium, 88039 Friedrichshafen, Germany}
\author{Tobias Ziegler}
\affiliation{EADS Astrium, 88039 Friedrichshafen, Germany}
\author{Indro Biswas}
\affiliation{German Aerospace Center, Pfaffenwaldring 38, 70569 Stuttgart, Germany}
\author{Christoph Seibel}
\affiliation{Universit\"at W\"urzburg, Experimentelle Physik und R\"ontgen Research Center for Complex Material Systems (RCCM), Am Hubland, 97074 W\"urzburg, Germany}
\affiliation{Gemeinschaftslabor f\"ur Nanoanalytik, Karlsruher Institut f\"ur Technologie (KIT), 76021 Karlsruhe, Germany}
\author{Mathias Schulze}
\affiliation{German Aerospace Center, Pfaffenwaldring 38, 70569 Stuttgart, Germany}
\author{Nico Brandt}
\affiliation{EADS Astrium, 88039 Friedrichshafen, Germany}
\author{Achim Sch\"oll}
\affiliation{Universit\"at W\"urzburg, Experimentelle Physik und R\"ontgen Research Center for Complex Material Systems (RCCM), Am Hubland, 97074 W\"urzburg, Germany}
\affiliation{Gemeinschaftslabor f\"ur Nanoanalytik, Karlsruher Institut f\"ur Technologie (KIT), 76021 Karlsruhe, Germany}
\author{Patrick Bergner}
\affiliation{EADS Astrium, 88039 Friedrichshafen, Germany}
\author{Friedrich T. Reinert}
\affiliation{Universit\"at W\"urzburg, Experimentelle Physik und R\"ontgen Research Center for Complex Material Systems (RCCM), Am Hubland, 97074 W\"urzburg, Germany}
\affiliation{Gemeinschaftslabor f\"ur Nanoanalytik, Karlsruher Institut f\"ur Technologie (KIT), 76021 Karlsruhe, Germany}

\begin{abstract}
The kinetic energy distributions of photo-electrons emitted from gold surfaces under illumination by UV-light close to the threshold are measured and analyzed. Samples are prepared as chemically clean through Ar-Ion sputtering and then exposed to atmosphere for variable durations before Quantum Yield measurements are performed after evacuation. During measurements the bias voltage applied to the sample is varied and the resulting emission current measured. Taking the derivative of the current-voltage curve yields the energy distribution which is found to closely resemble the distribution of total energies derived by DuBridge for emission from a free electron gas.
We investigate the dependence of distribution shape and width  on electrode geometry and contaminant substances adsorbed from the atmosphere, in particular to water and hydro-carbons.
Emission efficiency increases initially during air exposure before diminishing to zero on a timescale of several hours, whilst subsequent annealing of the sample restores emissivity. A model fit function, in good quantitative agreement with the measured data, is introduced which accounts for the experiment-specific electrode geometry and an energy dependent transmission coefficient. The impact of large patch potential fields from contact potential drops between sample and sample holder is investigated. The total quantum yield is split into bulk and surface contributions which are tested for their sensitivity to light incidence angle and polarization. Our results are directly applicable to model parameters for the contact-free discharge system onboard the LISA Pathfinder spacecraft.
\end{abstract}

\pacs{79.60.Bm,79.60.Dp,73.20.At}
\keywords{photoemission, photoelectron spectra, surface states, adsorbed layers}
\maketitle

\section{Introduction\label{Introduction}}
\subsection{Thematic Background: Photoemission }
Early experiments on photoemission (see e.g. \cite{Com13,Mil16}) were aimed to shed light on particular details of the emission process from various metals, such as the distribution and range of photo-electron energies. These experiments clearly showed a cutoff of the emitted electron energies at a maximum value, in agreement with the Einstein equation. It was not until the work of Fowler \cite{Fow31} that photo-electron emission was quantitatively connected to the underlying statistics of the free electron gas and temperature effects were shown to play a crucial role in smoothing out the threshold for emission \cite{DuB32}. Building upon the work of Fowler, DuBridge derived expressions for normal and total energy distributions of photo-electrons emitted close to the threshold \cite{DuB33} which were shortly afterwards confirmed by experiments \cite{DuB33-2,Roe33}.
\\General understanding of the theory of photo-emission was significantly enhanced in the 1960s and 1970s, when Spicer introduced the three-step model\cite{Spi58,Ber64}, where electrons are excited, migrate to the surface whilst undergoing inelastic collisions and finally escape over the potential barrier represented by the work function. Later, more refinements to the theory were made when investigations were made of surface related effects such as the modeling of surface states \cite{Hul85,Hof84} and surface plasmon excitations \cite{Dru88,Sut89}.
\\For this paper, DuBridge's work is a major focus as we build on his results and refine his model to account for different electrode geometries and the effects of adsorbates on the previously clean metal surfaces. We find that the original papers, which rely on a very simple model of the free electron gas based purely on the Fermi-Dirac statistics and without consideration of band-effects, are very well suited for describing our measurement data which were taken for emission close to the threshold, i.e. where the energy of the incident light is very close to the material work function $h\nu\approx\phi$. In a more recent paper it was shown that even if the perfectly free electron gas model which is equivalent with the "one orthogonalized plane wave (OPW) model" with dispersion relation $E+h\nu=\hbar^2p^2/2m$ is replaced by a more refined 2-OPW approximation, very similar results for electron escape function and consequently the photo-electric yield are obtained close to threshold\cite{Sha94}. Based on these findings and on the success of the three step model to describe general photo-electron emission experiments and many aspects of our experiment in particular, we legitimate the application of DuBridge's model for our analysis throughout the paper.
\subsection{LISA Pathfinder space mission}
Some major findings of this paper, notably the energy distribution, its width and the quantum yield, are used as model inputs for a detailed simulation to quantify stability and performance  of the discharge system\cite{Zie09} (and to aid the design of its onboard control algorithm) which is part of the scientific payload of the LISA Pathfinder mission\cite{Bel08}, a testbed for a gravitational wave detector in space. At the core of its working principle are two large test-masses which are monitored under free fall conditions and serve as ultra-sensitive inertial sensors. As these gold-coated test masses charge up inadvertently through cosmic radiation they must be discharged to mitigate adverse effects, which is done using a UV source via the photoelectric effect. The stability and efficiency of the discharge process is found to depend on the magnitude and variability of certain physical parameters, such as the quantum yield. The requirement for reproducibility of the latter and of other parameters is further aggravated by the need to expose the gold surfaces to ambient conditions during assembly and integration of the sensors.

\section{Experimental and Theoretical Background information}
\subsection{The Experimental setup}
\label{subsec_experiment}
\begin{figure}
 \includegraphics{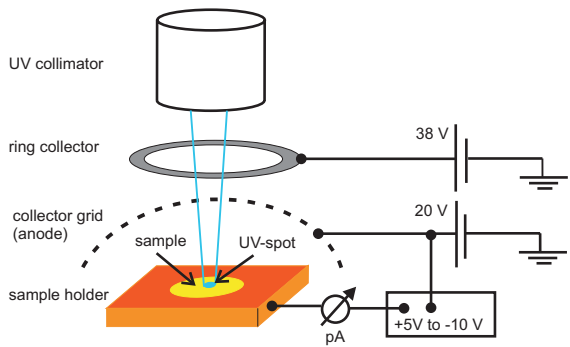}%
 \caption{\label{fig_1}A schematic of the experimental setup used for the measurements of the photo-emission currents at DLR.}
 \end{figure}

A schematic of our experimental scheme is shown in Figure \ref{fig_1}. Here we only present some essential details for the following discussions and refer the reader to reference \cite{Bis12} for more details.
The setup consists of a copper sample holder (4 x 6 cm) onto which the gold sample (typically circular with 2.5 cm diameter) is attached through two clamps, a field determining grid, a collector ring above the grid and a UV source with a collimating tube and a condenser lens.
All components except for the UV-source are installed inside a vacuum chamber with a typical pressure of $10^{-7}$ mbar.
The UV-light is emitted from a Hg-discharge lamp at the output of which wavelengths shorter than 230 nm are removed by a filter so that only light of wavelength   = 253.6 nm = 4.89 eV is transmitted. It enters the vacuum chamber through a fused-silica window and is subsequently collimated inside an aluminum tube. A single aspherical lens focuses the light to a spot size of 4.5 mm on the sample surface (see Fig. \ref{fig_1}). The UV intensity on the sample and the spot width were determined by means of a UVG100 diode (Int. Radiation Detectors).
The sample holder can be heated to temperatures above $130^{\circ}$ C in order to anneal the sample. A thermo-couple is attached for monitoring the temperature. The field determining collector grid is made of a fine mesh of Platinum which is curved in one plane (cylindrical) and mounted at a distance of approximately 3 cm to the sample center. Both electrodes are kept at a base potential of +20V with respect to the chamber housing, which is connected to the ground. The voltage of the sample holder with respect to the collecting grid was varied between +5 to -10 V during the emission current measurements. This voltage shall be referred to as bias voltage $V_b$ in the following. An additional ring shaped collector electrode at a constant potential of +38V is installed between aluminum tube and grid. The purpose of this electrode is to remove any electrons emitted from other parts of the setup than the sample and also to avoid that emitted electrons passing the collecting grid (covering approximately $50\%$ of the geometric area) may be reflected back towards the sample. This arrangement was necessary to obtain the correct emission current, which was recorded as the drain current utilizing a high precision electrometer.
Dielectric materials in the vicinity of the illuminated sample were found to considerably perturb the measurements due to electrostatic charging and caused a distorted field at the sample surface. In consequence, a supporting rod of alumina was covered with a metal tube, and Kapton insulated wires were moved well below the sample holder.

\subsection{Work Functions and Energy Distributions}
\begin{figure}
 \includegraphics[width=0.85 \linewidth]{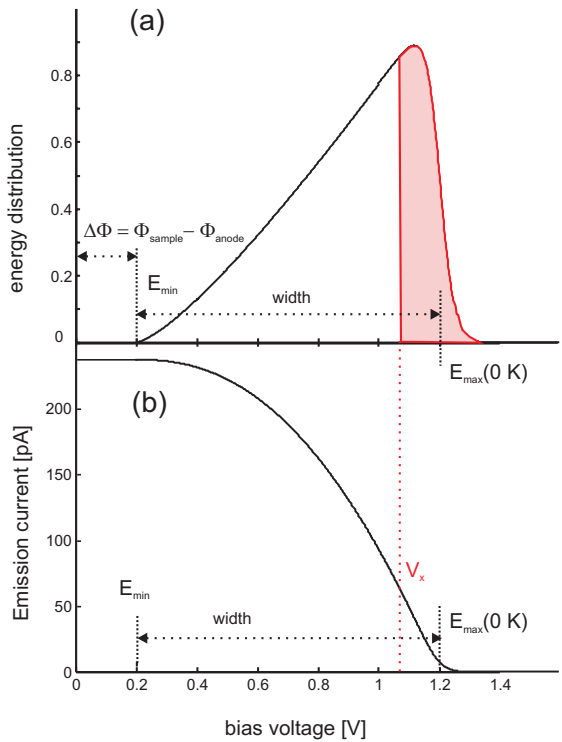}%
 \caption{\label{fig_2}The energy distribution (a) and current-voltage curve (b) are plotted against the applied bias voltage for a typical measurement. The minimum and maximum energies are indicated by dotted vertical lines where the former is offset from the origin by the contact potential and the latter is defined by the cutoff at zero temperature. For a given bias voltage all electrons with kinetic energies higher than $V_x$, indicated by the red shaded area underneath the curve in (a), overcome the bias field and contribute to the drain current, given by the intersection with the vertical dotted line in (b).}
 \end{figure}

The energies of the electrons emitted from the sample surface generally vary between 0 eV to a maximum energy given by the difference between the photon energy and the sample work function. However, the energetic position is shifted by the value of the contact potential, which is a consequence of the work function difference between sample and collector. This shift is identically found in the recorded distribution curves. The electrons are collected by the platinum anode and the emission current is measured for a set of different bias voltages which are incremented in small steps until the emission current has dropped to zero.
Based on these considerations the minimum and maximum energies of the measured energy distribution are given by \cite{Cah03}
\begin{eqnarray}
E_{\rm min}&=&\Delta\phi=(\phi_{\rm s}-\phi_{\rm a})\nonumber\\
E_{\rm max}&=&h\nu-\phi_{a},
\label{Equ::energies}
\end{eqnarray}
where $h\nu$ is the photon energy and $\phi_s$ and $\phi_a$ are the work functions of the sample and anode, respectively.
From Eqs. \ref{Equ::energies} we find that the sample work function can be determined from the distribution widths as follows \cite{Cah03}
\begin{equation}
\phi_s=h\nu-(E_{\rm max}-E_{\rm min})
\label{Equ::sampleWF}
\end{equation}
It is important to note that the characteristics of the receiving anode (work function, dipolar charge, etc...) are removed altogether when subtracting the distribution widths from the photon-energy as the difference only depends on the sample work-function.
\\Figure \ref{fig_2} (a) shows a typical distribution curve of total electron energies. Here we assumed that the electron energy is proportional to the applied bias voltage which will be discussed in more detail in section \ref{section_geometry}.
The sample bias voltage $V_b=V_{sample}-V_{anode}$ was incrementally increased and the corresponding Quantum Yield (QY) current, i.e. the number of electrons emitted from the sample surface per unit of time and per incident photon, was measured. It presents a barrier for the electrons to escape from the sample surface and therefore, with increasing bias, the number of escaping electrons and hence the anode current decrease. Only those electrons with higher kinetic energy than the bias voltage (including the contact potential) manage to escape. The measured drain current therefore represents an integration of the distribution function of kinetic energies from the bias voltage to infinity. This is indicated by the red-shaded area in Figure \ref{fig_2}. The voltages are given in units of the "stopping voltage" $V_m$ which is defined by the maximum energy of the emitted electrons at the sample surface: $eV_m=h\nu-\phi_s$. Figure \ref{fig_2}a shows that the energy distribution goes beyond the value of $E_{\rm max}$, which is an effect of the finite temperature of the distribution (300 K in our example). For zero temperature (T=0 K) the distribution would abruptly drop to zero at $E_{\rm max}$ and therefore define a sharp cutoff. We also note that the whole energy distribution is shifted by a positive value of $\Delta\phi$ which corresponds to the case where the work function of the sample is larger than the one of the collector.
\\Denoting the electron energy distribution curve as $p(E)$ one obtains the following relations between the distribution and the yield current \cite{DuB33}:
\begin{equation}
I(V_b)=\int^{\infty}_{V_b}p(V)dV\Rightarrow p(V)=-\frac{dI(V)}{dV}
\label{Equ::Den&IV}
\end{equation}
In the following sections we will repeatedly come back to the definition of Eq. \ref{Equ::Den&IV} when fitting the measured current-voltage curves and determining the underlying energy distribution functions.

\subsection{Theory of Fowler and DuBridge}
Fowler derived a theoretical expression for the expected photo-current as well as for the distribution of electron energies normal to the surface ("normal distribution"), valid for photo-emission from a free electron-gas close to the long wavelength emission threshold \cite{Fow31}. In the following we shall briefly recapture the basics of his derivation which will be useful for understanding later sections in this paper.
\\In a free electron gas the number of electrons per unit volume with velocity components in the ranges (u, u+du), (v,v+dv), (w,w+dw), where u is the component normal to the surface, is given - according to the Fermi-Dirac statistics- by:
\begin{equation}
n(u,v,w)dudvdw=2\left(\frac{m}{h}\right)^3\frac{dudvdw}{e^{\left[\frac{m}{2}\left(u^2+v^2+w^2\right)-\mu\right]/k_B T+1}},
\label{Equ::FermiDirac}
\end{equation}
where $k_B$ is Boltzmann's constant, $m$ the electron mass, $\mu$ the chemical potential, and $T$ the temperature. Integrating Eq. \ref{Equ::FermiDirac} over the velocity components parallel to the surface one finds the distribution of kinetic energies of the velocity components normal to the surface
\begin{equation}
n(u)du=\frac{4\pi k_B T}{m}\left(\frac{m}{h}\right)^3 \log\left[1+e^{\left(\mu-mu^2/2\right)/k_B T}\right]du
\label{Equ::normal_component}
\end{equation}
To find the number of electrons with "normal energies" in the interval given by $(E_n,E_n+dE_n)$  reaching a surface element of unit area per unit time, we multiply Eq. \ref{Equ::normal_component} by the normal velocity u and use the relation $E_n=mu^2/2$:
\begin{equation}
n(E_n)dE_n=\frac{4\pi k_B T}{h^3}\log\left[1+e^{\left(\mu-E_n\right)/k_B T}\right]dE_n
\label{Equ::normal_energy}
\end{equation}
Fowler then made the crucial ansatz that the number of electrons contributing to photo-emission scales proportional to the number of electrons for which the following condition is fulfilled
\begin{equation}
\frac{mu^2}{2}+h\nu=\mu+\phi_s
\end{equation}
This condition implies that the normal energy together with the energy of the absorbed photon suffices to overcome the potential step represented by the work function $\phi_s$ and lift the electron from the Fermi-level of energy $\mu$ to the vacuum level. We then assume that the yield current scales proportional to the integrated distribution function of the normal energies. Equation \ref{Equ::normal_energy} is then integrated over all normal energies and the integral simplified under the approximation that emission is close to the threshold so that $\mu\gg h\nu-\phi_s$.
Although there is no simple analytical expression the result can be approximated by a series expansion in the small parameter $\eta=(h\nu-\phi)/(k_B T)$ so that we find
\begin{equation}
I=C\frac{T^2}{\left(\mu+\phi-h\nu\right)^{1/2}} F(\eta)\label{Equ::Fowler_law},
\end{equation}
where $C$ is a constant that is independent from photon-energy, temperature, and work function and $F(\eta)$ is defined as follows:
\begin{eqnarray}
F(\eta)&=&e^{\eta}-\frac{e^{2\eta}}{2^2}+\frac{e^{3\eta}}{3^2}-...\hspace{10 pt}\eta\le0\nonumber\\
F(\eta)&=&\frac{\pi^2}{6}+\frac{\eta^2}{2}-\left(e^{-\eta}-\frac{e^{-2\eta}}{2^2}+...\right)\hspace{10 pt}\eta >0\label{Equ}\label{Equ::F_eta}
\end{eqnarray}
Taking the logarithm of Eq. \ref{Equ::Fowler_law} and assuming that the quantity $\left(\mu+\phi-h\nu\right)$ is approximately constant yields Fowler's law. It is based on the universal function represented by $\log F(\eta)$ and allows to determine the work function from the horizontal displacement of experimental data with respect to the universal function when $I/T^2$ is plotted against $\eta$ on a logarithmic scale.
\\Shortly after Fowler's seminal paper \cite{Fow31} DuBridge published a theoretical paper \cite{DuB33} discussing various aspects of measuring the distributions for normal (perpendicular to sample surface) and total energy.
The following high-precision measurements of the two distributions were in good quantitative agreement with DuBridge's predictions \cite{DuB33-2,Roe33}.
The derivation of the distribution of total energies is similar to the one for the normal energies and depends solely on the Fermi-Dirac statistics of the free electron gas.
The distribution for electrons within the interval $(v,v+dv)$ of the total velocity $v$ is given by:
\begin{equation}
N(v)dv\propto\frac{v^3dv}{e^{(mv^2/2-E_m)/k_B T}+1},
\label{Equ::energy_DB}
\end{equation}
where $E_m=eV_m=h\nu-\phi_s$ is the threshold energy.  If we now make the assumption that the total energy is used to overcome the external electrostatic potential , i.e. $m v^2/2=eV$, we obtain from Eq. \ref{Equ::energy_DB} the following expression for the total energy distribution as a function of applied bias voltage:
\begin{equation}
f_t(V)dV=A_t\frac{VdV}{e^{(V-V_m)e/k_B T}+1},
\label{Equ::E_total_V}
\end{equation}
where $A_t$ is a constant. This equation can in turn be integrated from the bias voltage to infinity (see Eq. \ref{Equ::Den&IV}) to obtain an expression for the expected current-voltage curve.
As was the case for the distribution of normal energies, there is no simple analytical expression for the integral but a Taylor expansion yields an approximate expression that we used in the fitting of the experimental data.

\subsection{The effect of geometry on the measurement}
\label{section_geometry}
\begin{figure}
 \includegraphics[width=0.85 \linewidth]{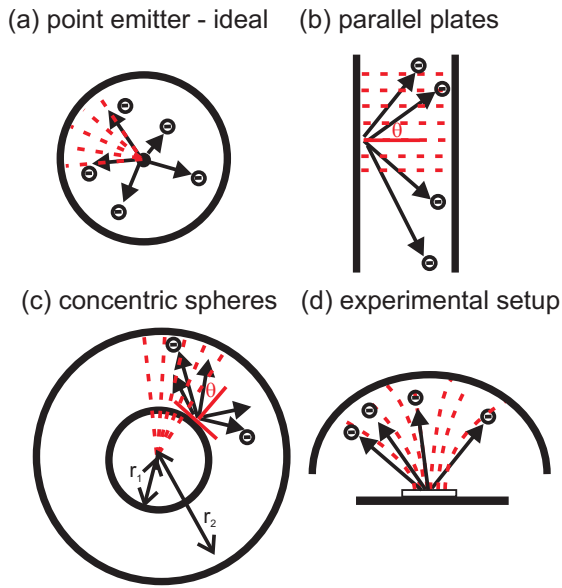}%
 \caption{\label{fig_3} Various electrode geometries are displayed for comparison. Electric field lines are indicated by the broken red lines, electron velocities by the black arrows. (a) In the ideal geometry for the measurement of total energy the emitter is a point source, all field lines are aligned with the electron trajectories. (b) In a parallel plate geometry the normal energies are measured. (c) The setup with two finite-sized concentric spheres lies in between the two extreme case a and case b. (d) The experimental setup of this paper resembles case c.  }
 \end{figure}

\begin{figure}
 \includegraphics[width=0.85 \linewidth]{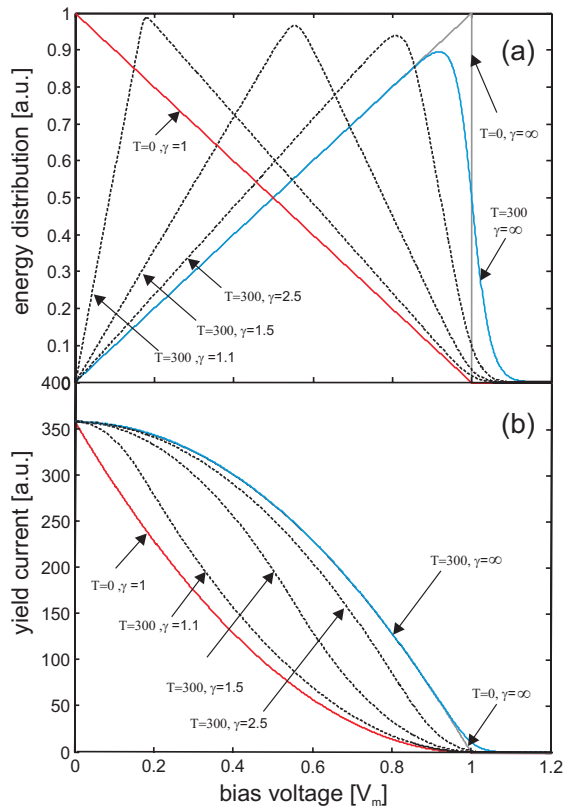}%
 \caption{\label{fig_4}The energy distributions of photo-electrons emitted from various electrode geometries are plotted in (a), the corresponding yield currents in (b). Voltages are given in units of the stopping Voltage $V_m$.
 The respective curves encompass geometries ranging from plane parallel ($\gamma=1$) to perfectly spherical ($\gamma=\infty$) and room temperature (T=300 K) is assumed in the curves represented by the black dotted lines. The distributions at T=0 K for the total energy (grey curve) and the normal energy (red curve) have a perfectly triangular shape. The total energy distribution for T=300 K is given by the blue curve.}
 \end{figure}

We already discussed that the electrode geometry plays an essential part in the actually measured distribution curves. The most important aspects to consider are the alignment of the electric field lines with respect to the electron trajectories and the solid angle extending from the emitting surface within which electrons may be captured by the collecting anode.  Various cases for electrode geometries are displayed in Fig, \ref{fig_3}.
\\{\bf 1.) Point emitter inside sphere}: In this setup the total energy distribution of the emitted electrons is measured.
The emitter is located in the center of the receiving sphere as shown in Fig. \ref{fig_3}a. All electrons move on trajectories which are perfectly parallel to the electric field lines so that the retarding fields decrease the total kinetic energy measured by the receiving sphere. This configuration closely resembles the early setups, where emitting samples were placed in the center of sealed evacuated glass spheres. The total energy distributions for T=300 K and T=0 K are given by the blue and grey curve in Fig.\ref{fig_4}, respectively.
\\{\bf 2.) Plane parallel plates}: In this setup, shown in Fig. \ref{fig_3}b, the distribution of normal energies of the emitted electrons is measured.
Here, the density of electric field lines remains constant between cathode and anode and the electron trajectories are not aligned with field lines.
In such a geometry the normal energy that is used to overcome the electrostatic potential is given by $E_n=eV$. We obtain from Eq. \ref{Equ::normal_energy} the following expression for the distribution of energies as a function of the applied bias voltage:
\begin{equation}
f_n(V)dV=A_n\log\left[e^{-(V-V_m)e/k_BT}+1\right]dV,
\label{Equ::En_V}
\end{equation}
where $A_n$ is a constant. The normal energy distribution for T=0 K is given by the red curve descending from the maximum at $V=0$ to zero at the stopping voltage $V=V_m$.
\\{\bf 3.) Two concentric spheres}: This setup is an interstage between the two extremes discussed before.
An arrangement of two concentric spheres, as the one shown in Fig. \ref{fig_3}c, does not accurately measure the total energy as the inner sphere is no point-like source but has a finite size. The electrons therefore follow trajectories which are quite generally not perfectly aligned with the field lines. If the radius of the emitting sphere approaches the one of the collecting sphere, we find that the geometry increasingly resembles that of two parallel plates, which was discussed in scenario (2) above. If, on the contrary, the radius of the inner sphere approaches 0, the case of a point emitter is approached, which was discussed in scenario (1) above. We therefore expect a smooth transition of the energy distribution function of scenario (3) when the radius of the inner sphere is changed from a value close to zero to a value close to the outer radius. We conclude that the concentric sphere measurement setup lies in between the two extremes of point emitter which measures the total energy and parallel plate emitter which measures the normal energy.
 The energy distribution which would be measured in a setup of two concentric spheres can be calculated by the convolution of the total energy distribution with the broadening function \cite{DiS70}. This has the effect of shifting as well as broadening the structure. The measured energy distribution for this geometry is described by a single electrode parameter, given by the ratio of the radius of the outer sphere to the radius of the inner sphere $\gamma=r_2/r_1$. From Fig. \ref{fig_4}a we see that by varying the geometry from $\gamma=\infty$ to $\gamma=1$ the energy distribution is gradually changing from the one for the total energy to the one for the normal energy, in agreement with our reasoning above. We should point out that these calculations rely on the assumption that the angular distribution of emitted photo-electrons follows a Lambertian distribution so that the probability $P(\theta)$ of emission at an angle $\theta$ with respect to the surface normal is given by $P(\theta)d\theta=\sin 2\theta d\theta$. It is interesting to note that -only based on a Lambertian distribution- the total energy distribution can be calculated from the normal one and vice versa, which clearly indicates the validity of the law. Considering that the original derivation of normal and total energy distributions by Fowler and DuBridge, respectively, was only based on arguments relating to statistics and energy conservation, we conclude that the very same arguments and approximations implicitly imply an angular dependency according to a Lambertian distribution. This conclusion is in agreement with a recent derivation and experimental of a Lambertian distribution for photo-emission from gold surfaces \cite{Pei02}.
\\{\bf 4.) The experimental setup for this paper}: Our setup is shown schematically in Fig. \ref{fig_3}d. The cylindrical shape resembles scenario (1) in one direction and scenario (2) in the other direction, and can be modeled very well by scenario (3) in the sense, that the density of field lines decreases with increasing distance from the emitting surface and the electron trajectories
are generally not aligned with the field lines. Based on the similarity to the case of two concentric spheres we adopted the latter model to describe the measured energy distributions in our setup. For an optimal fit of the measured energy distributions to a model function we left the parameter $\gamma$ free to vary and determined from the fits that $\gamma\approx 2.1$. This is in good agreement with our expectations based on a simple model where the sample holder (half-width 2 cm) is represented by a sphere of 2 cm radius and the collector grid (2-3 cm distant from sample holder) by a sphere of 4-5 cm radius, which should yield a geometric parameter $\gamma$ somewhere between 2 and 2.5. We also observed in our measurements that after increasing the distance between sample and grid (by approximately 1 cm) the parameter $\gamma$ increased slightly, which agrees well with model predictions and gives further evidence that the position of the measured distribution peaks is governed by the electrode geometry.

\section{Experimental Data and Analysis}
\subsection{The Measurement steps}
Photo-emission spectra of various gold-coated samples were investigated. The samples, produced by Selex Galileo, consist of a glass substrate covered with a 200 nm thick layer of Ti which is in turn covered by a 800 nm  gold layer. For some samples also Aluminum or Titanium were used as substrates.
Measurements were performed on a total of 6 samples and the following measurement steps were executed at the German Aerospace Center:
The sample was first Ar-Ion sputtered with an ion energy of 1 keV and a current of typically 5-6 µA for a period for a period of ~20 minutes. Sputtering was performed in a separate chamber (not the measurement chamber) and the sample was subsequently exposed to atmosphere when being removed from the sputter chamber. The sample was kept under atmosphere for a prolonged period of time, typically from several minutes to several hours or even days.
After exposure the sample was re-inserted into the measurement chamber which was then pumped down to a pressure of typically $10^{-7}$ mbar. Then the sample was left for a duration of several hours while basic yield measurements were performed intermittently. Finally, the sample was annealed at a temperature of ~$130^{\circ}$ C for a duration of 12h-24h before the last QY measurement was performed.
\\Similar steps were performed in W\"urzburg, where the samples were sputter-cleaned {\it in situ} without breaking the vacuum. Additionally, the gold samples could be covered with another coating (on top of the gold) in a preparation chamber which was equipped with a mass spectrometer. The main purpose of this second coating was to explore the choice of materials with higher emissivity than that of gold.

\subsection{Measured Distributions and Quantum Yields}
\label{section::distrib}
\begin{figure}
 \includegraphics[width=0.95 \linewidth]{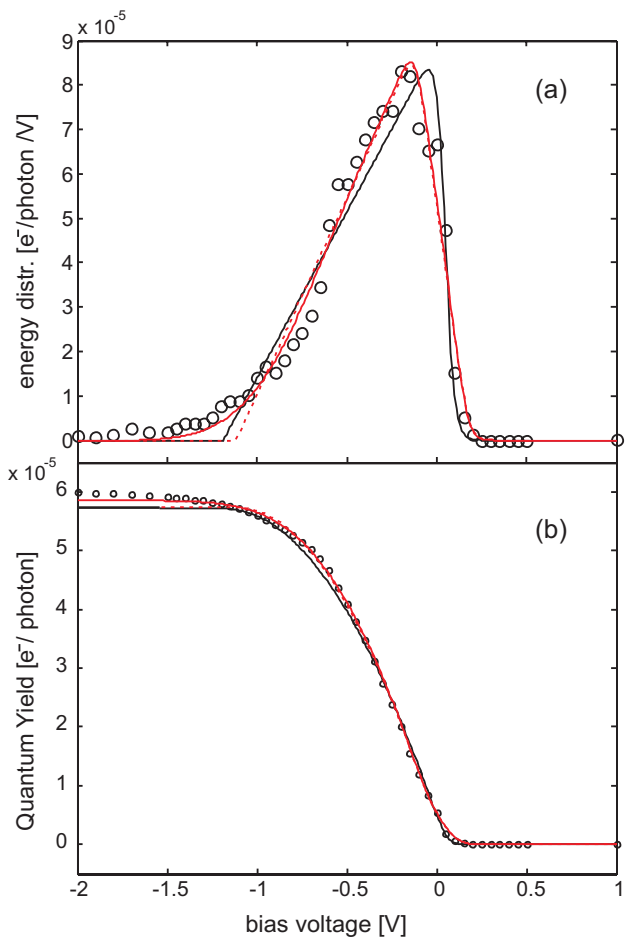}%
 \caption{\label{fig_5}The measured energy distribution (a) and current-voltage curves (b) for a given sample (ref.no. TP-04) after Ar-Ion sputtering, air exposure and annealing. The black circles denote the experimental data. The fitted energy distributions for an undetermined geometry ($\gamma$ free to vary) and for the total energy ($\gamma=\infty$) are given by the red dotted and black solid lines, respectively. The red solid line denotes a fit accounting for both, variable electrode geometry and a low-energy modification of the emission.}
 \end{figure}

\begin{figure}
 \includegraphics[width=0.95 \linewidth]{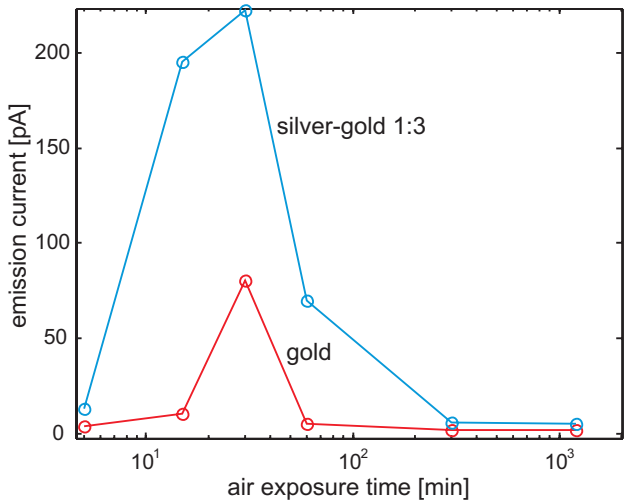}%
 \caption{\label{fig_6} The maximal emission currents are plotted against air exposure times for pure gold (red curve) and an Ag-Au mixture of 1:3 (blue curve).}
 \end{figure}

The data for a typical measurement performed after annealing are plotted as black circles in Fig. \ref{fig_5}, where the quantum yield is given in (b) and the derived energy distribution in (a). The black line represents a fit to the data based on an expression for the total energy (Eq.\ref{Equ::E_total_V} shifted by a Voltage offset $eV_0=\Delta \phi$), where the sample work function $\phi_{s}$, the distribution amplitude $A_t$, and the work function difference $\Delta\phi$ are the three free parameters. The dotted red line is obtained from a fit where the electrode geometry parameter may also vary and we find $\gamma=2.10$, in agreement with expectations. The two fits both yield values of $\phi_s\approx 3.7$ eV, $\Delta\phi\approx -1.2$ eV so that we obtain for the anode work function $\phi_a=\phi_s-\Delta\phi=4.8$ eV.
We should point out that all measurements performed during the campaign have yielded the similar values for $\phi_a$ which, in contrast to $\phi_s$, is not found to vary much between measurements. Complimentary fits to the high energy tails of the distribution, based on Fowler's function of Eq. \ref{Equ::F_eta}, yielded the same results for $\phi_a$ within 50 meV.
\\When comparing the solid black to the dotted red line we find that the latter curve fits the data much better with the corresponding least-square residuals being 30\% lower than those of the former curve. This is owing to the fact that our electrode geometry does not resemble a point-emitter inside a sphere and therefore does not measure the total energy distribution. However, neither curve gives a good fit for the low-energy tail of the data, where the yield current keeps increasing slightly with decreasing bias voltages down to -1.5 V. Furthermore, the measured shape of the energy distribution is slightly curved for lower energies whereas the theoretical expression for the total energy and the geometry-related derivatives thereof (see Fig.\ref{fig_4}a) clearly have a linear slope towards lower bias voltages.
\\This observation is in agreement with the early findings of Hill for emission from sodium \cite{Hil38}. An explanation was given in terms of an energy-dependent transmission coefficient describing the barrier the electron needs to overcome in the emission process.
From the historic debate it is well known that an image force (hyperbolic) potential yields a constant transmission coefficient \cite{Nor28} which is in disagreement with measurements of the thermionic emission constant and cannot explain other emission-related properties either. Neither does a simple square potential barrier provide a satisfactory description of the transmission barrier \cite{Con31}. There is still some debate on the exact shape of the latter but it is commonly agreed that intra-electron exchange forces play a dominant role. Charge separation and adsorbates at the material surface have a major impact on shape, depth, and associated work function of the potential. For this paper we exposed the gold samples to ambient air which comprises numerous possible contaminant molecules for the gold surface. In particular, these include water and hydro-carbons whereas oxygen has negligible adsorption probability on gold\cite{Fai74}.
\\Numerous experiments have been performed on the adsorption of water on various metals and semiconductors\cite{Hen02}, where this has generally led to a decrease of the work function between a few hundred meV to more than 1 eV, but in some instances also the opposite was true, depending on the orientation and packing densities of the water molecules in the adsorbed layer. Water also has the general tendency to cluster on metal surfaces, which is particularly true for noble metals such as gold where the surface interaction with water is weak and water-water interactions are significant in comparison \cite{Hen02}. Electron transmission through films of hydrocarbon chains and organized organic thin films have also been investigated over the past years, which was generally done using low-energy electron transmission (LEET) spectroscopy but recently also by studying the ballistic motion of photo-electrons \cite{Kad95}.
\\An easy to measure parameter is the asymptotic Quantum Yield (QY) which is obtained for large negative bias voltages, ensuring that all electrons which are emitted are actually counted towards the total yield. This parameter is used to compare the total emissivity between samples for different air exposure times. In this respect we made some interesting observations. Directly after Ar-Ion sputtering and without previous exposure to atmosphere the gold coated samples were found to be non-emissive (yield current smaller than detection limit of $\approx 1$ pA). When the samples were exposed to ambient air for a controlled period in time before the subsequent measurement under evacuated conditions, the yield was found to increase for a timescale on the order of several minutes before starting to drop again. Finally, the yield reached zero after an exposure time somewhere between one and several hours and remained at zero for exposure times up to several days. This was independently confirmed by different measurements in W\"urzburg and at DLR.
The typical emission currents for a gold coated sample and another sample coated with an Au/Ag 1:3 mixture are given in Fig.\ref{fig_6}. In this respect we point out that systematic measurements of the work function $\phi_s$ of Au(111) and of Ag thin films deposited on Au(111) substrates have recently been performed using AES and ARPES\cite{Cer06}, where a major focus has been to characterize intermixing and alloy formation at the surface as well as the properties of surface states.
\\When comparing the measurements for different samples we found that there was generally a large variation of the duration when emission peaked and had dropped to zero again. This is most likely attributable to slightly varying ambient conditions, in particular humidity, when the samples were exposed to air in the laboratory. Based on our measurement results we believe that the observed phenomena are likely due to co-adsorption of water together with hydrocarbons onto the gold surface. Further indications that this might be true are obtained from the results of XPS measurements of the contaminated surfaces where signals of carbon and oxygen alone were found in the emission spectra. Water usually adsorbs and desorbs on very short timescales at room temperature \cite{Hen02}, however the co-adsorption of water with other contaminants could tilt the balance between adsorption and desorption towards the former process and impart a certain stability to the adsorbed surface layers.
As the vacuum chamber in which all measurements were performed in W\"urzburg was equipped with a non-collimated Al X-ray source (1486 eV) in addition to the mercury discharge lamp (4.89 eV), such XPS as well as the photo drain current data could be collected in situ.
\\After annealing the samples at a temperature of approximately $130^{\circ}$ C for a duration between 12h to 24 h we found that we could reproducibly restore high emissivity for each sample. The quantum yield after annealing had an average of QY=$7.4\times 10^{-5}$ electrons per incident photon (at perpendicular incidence)  with only small differences of around 20-30$\%$ between samples. The observation that the yields were very similar after annealing for all samples irrespective of previous differences in exposure times was very encouraging and crucial for restoring equal emissivity to all samples. The latter property is essential for the proper functioning of the contactless discharge subsystem for the inertial sensor subsystem of the LISA Pathfinder spacecraft \cite{Bel08}.

\subsection{Modeling the Transmission coefficient}
\begin{figure}
 \includegraphics[width=0.95 \linewidth]{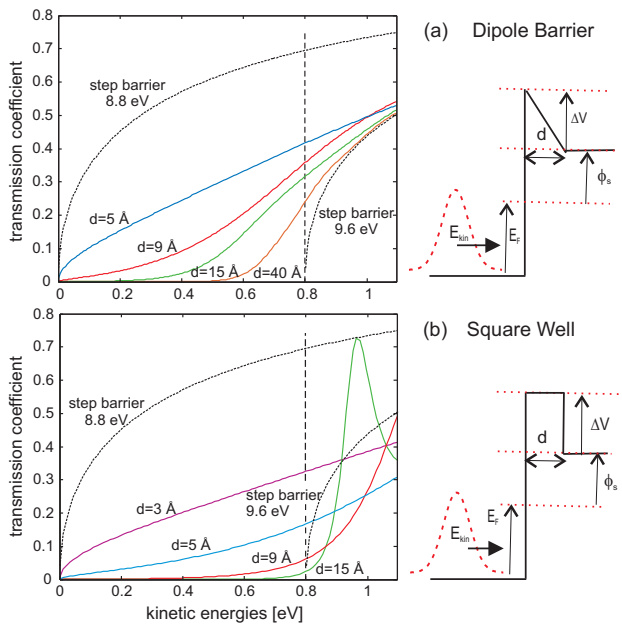}%
 \caption{\label{fig_7}The transmission coefficients are plotted for (a) a dipole potential barrier and (b) a square potential barrier on top of a square well for various barrier thicknesses between 2 and 40 {\AA}. The barrier height, as defined in the schematics on the right side, is chosen as $\Delta$V=0.8 eV in both cases. Electron energies are referenced to $\phi$=3.8 eV and the Fermi energy is taken to be $E_F=5$ eV. Transmission coefficients for basic square wells of height $E_F+\phi$ and $E_F+\phi+\Delta V$ are given by the dotted lines for comparison. The dashed vertical line denotes the point where the electron energy equals the maximum of the potential.}
 \end{figure}

Whereas an accurate model of electron transmission through layers or clusters of water and hydrocarbons requires a three-dimensional model,  in this paper only two basic one-dimensional models are used for illustration purposes. The transmission curves for electron motion through a triangular (dipolar) and a square potential on top of a square well are shown in Fig.\ref{fig_7}a and b, respectively, for a set of layer thicknesses, each layer having a potential height of $\Delta$ V=0.8 eV. For comparison, we also plotted the transmission curve for a square well alone (black dotted curve) and a second square well of added height $\Delta$V. For increasing layer thickness the transmission coefficients change from a curve resembling transmission through the lower square well (8.8 eV) to one describing transmission through the higher potential well (9.6 eV). We would like to point out that there is a pronounced similarity between the curves shown in Fig.\ref{fig_7}a and those measured in Ref.\cite{Kad95} for transmission through dipolar organic films.
Transmission through a square potential on top of a square well, shown in Fig. \ref{fig_7}b, looks somewhat different compared to the dipolar case in (a). Most notably, we observe no asymptotic approach to the transmission curve of the deeper square well but rather the occurrence of resonances for larger layer thicknesses (see green curve).
\\Our experimental data in Fig.\ref{fig_5}a are well fitted (red solid line) by a modified distribution which is obtained from the geometrically adjusted one (red dotted curve) by multiplication with an energy dependent transmission coefficient corresponding to a dipolar barrier of 8.4 ${\rm \AA}$ thickness. These are typical parameters for a multi-layer water film on a metal surface (see simulations in \cite{Nam98}).
A reasonable fit, not quite as good as for the dipolar barrier (larger residuals), is obtained for a transmission coefficient that vanishes linearly with diminishing energy, such as the one assumed from early experiments by Houston \cite{Hou37} and further investigated by Buckingham \cite{Buc50}.
It should be stressed at this point that this is a simplified model and the low energy tails in Fig. \ref{fig_5} more likely have a different origin. We shall investigate several other plausible explanations, in particular electrostatic disturbances, in the following sections.

\subsection{Investigating the Distribution Widths}
\begin{figure}
 \includegraphics[width=0.95 \linewidth]{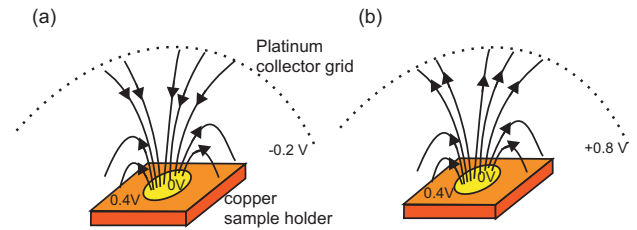}%
 \caption{\label{fig_8} The force lines between gold sample surface, copper sample holder and collecting mesh are plotted for two different bias voltages, a positive bias voltage in (a) and a negative one in (b).}
 \end{figure}

\begin{figure}
 \includegraphics[width=0.95 \linewidth]{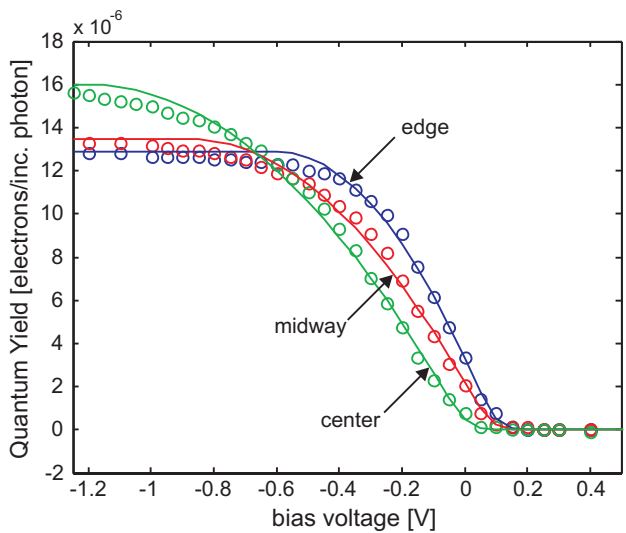}%
 \caption{\label{fig_9} Plot of the measured yield curves when the position of the UV-spot is moved from the sample edge (blue curve), through a midway point (red curve) to the sample center (green curve). The distribution widths increase and the minimum energies shift towards more negative bias voltages whilst the maximum energies remain nearly unaffected.}
 \end{figure}

From our measurements we find that the average distribution width is given by $\Delta E=1.2$ eV , implying effective work functions (see Eq.\ref{Equ::sampleWF}) of 3.7 eV which is significantly lower than the typically quoted values of 4.2-5.1 eV \cite{And59,Som68} for clean gold and of 4.2 eV for gold after air exposure\cite {Feu72}. There is good agreement, however, with previously published results of air exposed gold films on top of a tantalum substrate which were vacuum-sealed in a glass tube \cite{Ran80}. Our results were reproduced in a series of measurements performed on different samples over a prolonged period of approximately 2 months. During this campaign it was necessary to open the experimental chamber and break the vacuum, insert a new sample, and finally evacuate the chamber again before each measurement. After the initial bake-out at the beginning of the campaign no more chamber bake-out was performed in between measurements and we believe this to be the reason why we observed an increase of the distribution widths with time. In between measurements the vacuum chamber is increasingly contaminated by atmospheric compounds, in particular water, which adsorbs to the chamber walls under ambient conditions and may desorb from there to contaminate the sample during measurements. Patch fields created by water clusters on the sample but also those on the sample holder may further contribute to a broadening of the curves. To test this hypothesis we performed another bake-out of the vacuum chamber during 5 days at a temperature of approximately $100^{\circ}$ C. Subsequent photoemission  measurements found that the distribution widths were around 1.1 eV and therefore clearly smaller than before bake-out, the values being similar to the first measurements after the initial bake-out. Water peaks were also prominently visible in mass spectrometer measurements taken in the vacuum chamber to investigate the composition of its residual gas. Furthermore, XPS measurements gave no indication of other contaminants than carbon and oxygen on the sample surface (hydrogen being undetectable to XPS).
The broadness of $>3$ eV of the oxygen 1s signal at 531 eV hints at the presence of organically bound oxygen as well as water-bound oxygen or oxides. These observations seem to indicate that the increase in curve widths is probably attributable to the presence of water.
\\However, there are other prominent effects which may affect an increase in distribution width and therefore contribute to the overall uncertainty of the absolute width and, accordingly, of the sample work function:
On the one hand, we observed a strong sensitivity of the measured energy distribution curves to the presence of any dielectric component in the vicinity of the emitting sample (see discussion in section \ref{subsec_experiment}) which tended to broaden the curves. Similarly, small dielectric zones on the sample itself (where the gold and titanium coatings had peeled off in a tiny fragment) lead to an observable broadening of the curves. We therefore made very effort to ensure all such components were either replaced, shielded or removed from the field of view.
\\On the other hand, the work function difference between the copper sample holder and the attached gold sample (typically  $\phi_{Cu}$=4.7 eV and $\phi_{Au}$=5.1 eV for clean metals)\cite{Som68} affects a contact potential drop between the two. The higher work function of the gold leads to a flow of electrons from copper to gold and therefore to an increased potential of the copper surface with respect to the gold surface. The potential drop essentially presents a large patch field around the gold sample which affects the field lines and therefore modifies the (kinetic) energy of the emitted electrons. This is schematically depicted in Fig. \ref{fig_8}.
In some dedicated measurements we aimed to explore this effect by changing the location of the UV spot with respect to the center of the gold sample and studying the effect on the measured distribution width. The UV beam was first positioned at the edge of the gold sample ($\sim 1$ cm from the center), then midway between the edge and the center, and finally at the center of the gold sample. The measured distribution widths were 0.69 eV, 0.95 eV, and 1.20 eV, respectively, implying that close to the sample edge the measured energy distributions are significantly narrower than at the sample center. At the same time we observe that the minimum energy $E_{min}$ of the distribution (see Eq. \ref{Equ::energies}) assumes values of -0.6 eV, -0.9 eV and -1.2 eV for the three positions, respectively, whereas the maximum energy $E_{max}$ remains nearly constant.
The measured yield curves are given in Fig.\ref{fig_9}.
\\This effect can be qualitatively understood from Fig. \ref{fig_8}. Without loss of generality  we shall ignore the contact potential difference between sample and collecting anode for the following discussion as this merely constitutes a voltage offset which we absorb into the general definition of the bias voltage. In this picture $\Delta\phi=0$ and photoemission starts at a large positive bias voltage (cf. Fig. \ref{fig_8} a) equal to the sample work function whilst it ends at zero bias voltage, where electrons with low kinetic energies are emitted. When measuring the emission threshold around $E_{max}$ we apply comparatively large positive bias voltages and therefore strong fields between the electrodes so that the effect of the copper patch field, which is highly nonlinear and restricted to a small region around the sample edge, is rather small.  If, however, we measure the yield current saturation around $E_{min}$ we apply very small or no external fields so that the copper patch field becomes very significant and shifts the saturation point from zero towards positive bias voltages. The yield curves in Fig. \ref{fig_9} clearly show this behavior. When the UV light spot is positioned in the sample center the general emission distribution width is least affected whereas at the sample edge, where patch fields are much stronger, the distribution width is significantly decreased. The curves were fitted with an approximate analytic expression for the integrated distribution of the total energy (Eq. \ref{Equ::E_total_V}) to extract the widths and minimum energies. The fits of the respective yield curves are given by the solid lines.
To further investigate the copper sample holder we performed a measurement with a very large square sample of a thin gold layer on top of an aluminum substrate which covered nearly the whole area of the sample holder. The measured distribution width of $\sim 1.1$ eV agrees well with the measured width in the center of the smaller sample based on a circular glass substrate. From these observations we concluded that the general distribution widths are only insignificantly affected by the patch fields of the sample holder as long as the UV beam is well centered on the smaller circular samples. However, even then it seems quite possible that the patch fields could have some impact on the slope of the distribution towards low energies (cf. section \ref{section::distrib}).
\\Finally, from electrostatic calculations we concluded that the applied potential of $\sim 20 V$ between the anode grid and the chamber walls leads to distortions of the electric field between the electrodes, as this volume is not fully shielded by the anode grid against perturbations from the chamber walls. Numerical simulations indicate that this may lead to a significant broadening of the measured distribution widths. Mitigating measures could be to reduce the potential difference and a more comprehensive enclosure of the sample surface by the anode mesh.

\subsection{The Surface Effect}
\begin{figure}
 \includegraphics[width=0.95 \linewidth]{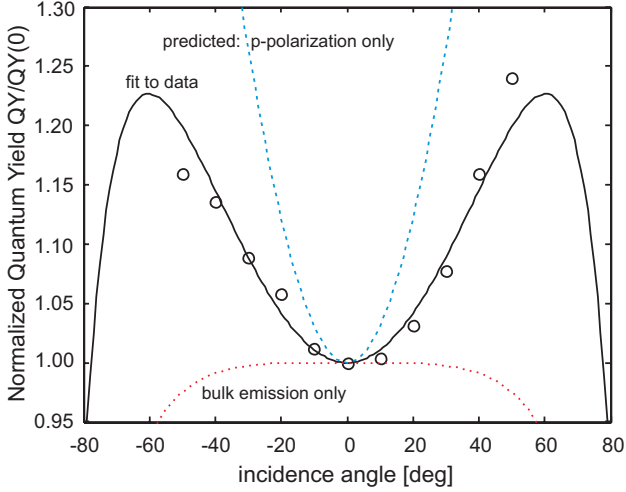}%
 \caption{\label{fig_10} The measured quantum yield (open circles) is plotted against the light incidence angle. The black solid line denotes a fit using Eq. \ref{Equ::Surface} for the vectorial photo-effect and the dotted red line marks the expected yield for bulk emission only. The blue broken line is the inferred yield for purely p-polarized light. The measurements are normalized by the yield value at normal incidence.}
 \end{figure}

Our analysis so far has dealt with the so-called bulk properties of the emission process, where surface properties such as adsorbed molecules and surface dipoles only play an indirect role through their impact on the transmission barrier and material work function. However, apart from the contribution of bulk emission to the total yield, there is another significant contribution originating from surface field mediated emission, the vectorial photo-effect\cite{Bro71}. It scales proportional to the electromagnetic energy inside the solid which is associated with the electric field vector that is normal to the surface and can be easily separated from the bulk contribution due to its dependence on the incidence angle and polarization of the light. The two effects, bulk- and  vectorial photoemission, vary with incidence angle in a completely different way \cite{Ped08}.
\\Bulk emission varies in proportion to the power of the absorbed light which depends linearly on the absorption coefficient: $QY_{bulk}\propto (1-R(\theta))$, where $R(\theta)$ is the sample reflectivity as a function of the incidence angle $\theta$. However, a general expression for the quantum yield, specifically including surface emission, is expected to scale as \cite{Bro71}
\begin{equation}
\frac{QY(\theta)}{QY(0)}=\frac{\epsilon_s(\theta)+\epsilon_{p\parallel}(\theta)}
{\epsilon_s(0)+\epsilon_{p\parallel}(0)}+r\frac{\epsilon_{p\perp}(\theta)}{\epsilon_s(0)+\epsilon_{p\parallel}(0)},
\label{Equ::Surface}
\end{equation}
where $\epsilon_s$, $\epsilon_{p\parallel}$, and $\epsilon_{p\perp}$ are the electromagnetic energies inside the solid due to electric field components associated with s-polarization, p-polarization parallel to the surface, and p-polarization normal to the surface, respectively. These quantities can be calculated directly from the Fresnel equations in combination with expressions for the field amplitudes transmitted through the solid surface\cite{Bro71}. The parameter 'r' is a measure of the photoemission efficiency of perpendicular fields compared to parallel fields, and a value of r=1 indicates bulk emission which is simply proportional to the total amount of absorbed light. A series of measurements was performed where the angle of the incident light was varied in steps of 10 degrees from normal incidence up to an incidence angle of 50 degrees on either side of the surface normal. The data are given as the open circles in Fig. \ref{fig_10}, which are fitted well by Eq.\ref{Equ::Surface} for r=4.67 (black solid line), indicating that emission through fields perpendicular to the surface is nearly five times as efficient as emission through parallel fields.
Note that the yields are normalized with respect to the yield obtained for normal incidence. For comparison, the expected behavior of bulk emission alone is given by the dotted red line. Considering that our light source is unpolarized we used the values of mixed light for the reflection coefficient in the scaling law for bulk emission to obtain this curve. Clearly, the predicted trend of bulk emission points in the opposite direction to the actual observations and to the fit.
\\We conclude that surface effects play a significant role in the emission process, increasing the total yield by more than 20$\%$ over a 50 degree variation of the incidence angle. The predicted behavior for p-polarized light is given by the dashed blue line which is calculated from Eq. \ref{Equ::Surface} by setting $\epsilon_s=0$.
Interestingly, we observed a disappearance of the sensitivity to the incidence angle when the sample was illuminated for a long time by the UV-source before the data were taken. This points towards a UV surface cleaning effect \cite{Vig86} of the illuminated region and the disappearance of the adsorbed surface layer which seems to have an impact on the vectorial photoeffect.

\section*{Conclusions}
We have measured and analyzed the specifics of photoemission from gold-coated substrates which had previously been exposed to atmosphere under illumination by ultraviolet light close to the threshold.
Important findings are that the distribution of emitted electrons is well described by the DuBridge expression for the total energy \cite{DuB33}, that the quantum yield is similar for all samples which were annealed after exposure ($\approx 7\times 10^{-5}$) and that the distribution widths, which may be affected by various electrostatic perturbations, are about $\approx 1.2$ eV, in good agreement with a previously published result \cite{Ran80}. We have also shown how the measured electron kinetic energy distributions change with the electrode geometry and applied a quantitative calculation method to convert distributions from one geometry to another, in particular to the geometry of our experimental setup. The impact of dipolar adsorbates is also discussed with respect to work-function, yield and distribution widths. After prolonged exposure to atmosphere we have found that the sample yield initially increases before dropping to zero on a timescale of hours, but emissivity can be restored through annealing.
Patch fields, introduced by contact potential drops between the sample and its holder, have likewise been investigated experimentally and good qualitative agreement has been found with theoretical models.
We have also shown how the total quantum yield can be split into a contribution from the bulk and one from the surface. The latter depends on the component of the electromagnetic field that is perpendicular to the surface and therefore strongly varies with the light incidence angle.
\section*{Acknowledgements}
The measurements and analyses described in this paper were performed in the frame of a contract with the European Space Agency (ESA) within the LISA Pathfinder project and the authors gratefully acknowledge the financial support. We would like to thank a number of people who have contributed to our progress in understanding and supported the experimental activities, in particular: Felix Erfurth and Markus Pfeil (TWT GmbH) for their invaluable support in the experimental preparations, Giovanni Taglioni (Selex Galileo) for sample fabrication, Luca Pasquali (University of Modena), Stefano Nannarone (BEAR Trieste) for their expert advice and support, Patrick Bergner, R\"udiger Gerndt and Ulrich Johann (Astrium), Gerhard Heinzel (Albert Einstein Institut), Bengt Johlander (ESA), Stefano Vitale and Rita Dolesi (University of Trento), Peter Wass and Tim Sumner (Imperial College London) for their advice and many fruitful discussions. We would also like to thank C\'{e}sar Garc\'{\i}a Marirrodriga and Giuseppe Racca (ESA) for their support and encouragement for these activities.

%

\end{document}